\documentclass[fleqn,usenatbib]{mnras}

\usepackage{newtxtext,newtxmath}

\usepackage[T1]{fontenc}

\DeclareRobustCommand{\VAN}[3]{#2}
\let\VANthebibliography\thebibliography
\def\thebibliography{\DeclareRobustCommand{\VAN}[3]{##3}\VANthebibliography}

\usepackage{graphicx}	
\usepackage{amsmath}	

\title[Tensions and signatures of new physics]{On dataset tensions and signatures of new cosmological physics}

\author[Marina Cort\^{e}s and Andrew R.~Liddle]{Marina Cort\^{e}s and Andrew R.~Liddle\\ Instituto de Astrof\'{\i}sica e Ci\^{e}ncias do Espa\c{c}o, Faculdade de Ci\^{e}ncias, Universidade de Lisboa, 1769-016 Lisboa, Portugal}

\date{\today}

\pubyear{2024}

\begin{document}
\label{firstpage}
\pagerange{\pageref{firstpage}--\pageref{lastpage}}
\maketitle

\begin{abstract}
Can new cosmic physics be uncovered through tensions amongst datasets? Tensions in parameter determinations amongst different types of cosmological observation, especially the `Hubble tension' between probes of the expansion rate, have been invoked as possible indicators of new physics, requiring extension of the $\Lambda$CDM paradigm to resolve. Within a fully Bayesian framework, we show that the standard tension metric gives only part of the updating of model probabilities, supplying a data co-dependence term that must be combined with the Bayes factors of individual datasets. This shows that, on its own, a reduction of dataset tension under an extension to $\Lambda$CDM is insufficient to demonstrate that the extended model is favoured. Any analysis that claims evidence for new physics {\it solely} on the basis of alleviating dataset tensions should be considered incomplete and suspect. We describe the implications of our results for the interpretation of the Hubble tension.
\end{abstract}




%
\section{Introduction}
One of the driving forces in current cosmology is the existence of tensions amongst various datasets, where fits appear to point to different values of one or more parameters. Most prominent are the strong Hubble tension between early- and late-Universe probes of the expansion rate (see \cite{VTR}, \cite{tensionrev}, \cite{SLL}, \cite{KamRiess} and references therein), and the weaker mismatch in power spectrum amplitude between cosmic microwave background (CMB) and weak-lensing observations \citep{Heymans,Amon,Secco,DESKiDS}. Possible explanations include statistical fluke, omitted or underreported systematic errors, mistakes/bugs in data analysis pipelines, or inadequacy of the physical model being fit to data.

To quantify the level of tension between two datasets $D_A$ and $D_B$ (which for simplicity only we will take to be fully independent) interpreted under model $M_1$, a popular tension metric is the Bayesian ratio $R^{AB}_1$ introduced by \cite{MRS}, defined by
\begin{equation}
\label{e:rdef}
R^{AB}_1 \equiv \frac{P(D_A,D_B|M_1)}{P(D_A|M_1) P(D_B|M_1)} \,.
\end{equation}
This compares the likelihoods under two assumptions: the numerator requires the two datasets to be jointly fit by a single set of parameter values, while the denominator permits (within the same overall model) each dataset to be fit by different parameter values, as if each dataset lived in a distinct universe. 

It is vital throughout to keep in mind that any dataset tension is associated not just to the datasets but also to the model(s) being assumed, so we always note these dependencies explicitly, datasets by letters and models by numbers. The statement about distinct universes is a little hard to interpret, but clearly relates to the extent to which the different datasets disagree on the parameter values. However, as several authors have pointed out, Bayes theorem immediately lets it be rewritten as \citep{March,HL1,Lemos}
\begin{equation}
R^{AB}_1 = \frac{P(D_A|D_B,M_1)}{P(D_A|M_1)}  \quad \left[ = \frac{P(D_B|D_A,M_1)}{P(D_B|M_1)}\right] \,.
\end{equation}
This quantifies, under a particular model assumption $M_1$, whether the existence of dataset $D_B$ makes dataset $D_A$ more or less probable than if dataset $D_B$ didn't exist. This now sounds like a very natural formulation of what one would mean by datasets being in tension. It is also nicely symmetric under exchange of datasets. Hence this tool has been widely utilised.

A variety of other tension metrics have been introduced and applied. These include the Index of Inconsistency \citep{linishak}, Bayesian Suspiciousness \citep{HL2}, parameter differences \citep{Raveri,RZH}, and the eigentension \citep{Park}. Each has its own subtleties and potential advantages and disadvantages, as summarised by \cite{Lemos}. But they lack the axiomatic simplicity and unambiguity of the Bayesian approach and so for simplicity we restrict our discussion to the Bayesian example. Nevertheless the same issues will broadly arise whatever tension metric is used.

The Bayes tension ratio was introduced to assess the compatibility of two datasets, its use exemplified by the Dark Energy Survey (DES) Y1 analysis \citep{DESyr1} where the ratio was required to exceed a certain threshold before the datasets were deemed to be combinable (again all within the context of whichever model has been chosen to explain the datasets). In this view, the tension ratio alerts us to the possibility of dataset incompatibility, though without telling us its origin or selecting a culprit. 

However, the tension ratio has since been co-opted to a different purpose, which is to provide support for one model over another. The idea is that if the dataset tension is less under a different model assumption $M_2$ than it is under $M_1$, this supports $M_2$ as a better description of the combined data. A typical statement is of the type `Our new model reduced the Hubble tension from 4.3$\sigma$ to 2.4$\sigma$ so our model is favoured over $\Lambda$CDM' (e.g.\ Table I of \cite{olympics}). If this new model features additional parameters, these will be associated to the discovery of new physical processes relevant to the datasets. Our purpose here is to challenge the usefulness, and correctness, of this view.

\section{Tension is only part of model probability updating}

\subsection{Bayesian model probability updating}

To use tension as an indicator of new physics, the working hypothesis now is that both datasets are correct, something that the tension metric was originally intended to diagnose rather than assume. The rest of this article will operate under that assumption, though at the end of an analysis seeking new physics one should reconsider whether the tensions might have other or multiple causes.

As everything is already within a Bayesian framework, the question of which model is a better description is uniquely answered by the posterior model probability ratio $P(M_1|D_A,D_B)/P(M_2|D_A,D_B)$ after application of all available data. Via Bayes' theorem this is related to the prior model probability ratio as
\begin{equation}
\frac{P(M_1|D_A,D_B)}{P(M_2|D_A,D_B)} = B_{12}^{AB} \frac{P(M_1)}{P(M_2)} \,,
\end{equation}
where the Bayes' factor
\begin{equation}
\label{e:bf}
B^{AB}_{12} = \frac{P(D_A,D_B|M_1)}{P(D_A,D_B|M_2)}
\end{equation}
is the ratio of model likelihoods and $P(M_1)$ and $P(M_2)$ are the prior model probabilities. The model likelihoods, also known as the Bayesian evidence, are the likelihoods averaged over each model's prior parameter space. Nested sampling \citep{Skilling} is a popular algorithm for carrying this out, cosmological implementations including CosmoNest \citep{CosmoNest}, MultiNest \citep{MultiNest}, and PolyChord \citep{PolyChord}. By averaging over the prior parameter space, the Bayesian evidence rewards overall goodness of fit and penalizes models with less predictiveness for the data in question. Care must be taken in consideration of the impact of noise on the statistic \citep{JP,Joachimi}.

As usual in a Bayesian calculation everyone is entitled to their own opinion on the prior model probability ratio,\footnote{As well as the parameter priors which we are assuming have been agreed on in advance.} whose freedom then propagates to the posterior ratio. But everyone will agree on the Bayes' factor and hence on which model has been favoured by any newly-incorporated data. The Bayes' factor is somewhat analogous to a derivative of the model comparison -- it indicates the direction in which the new data has taken the conclusion, saying for example that $M_2$ has become ten times more probable relative to $M_1$ as a result of new data, regardless of how likely you might have considered it to be beforehand or indeed whether there are other as-yet-unconsidered models.

\subsection{Tension metric ratios}

Taking reduction of tension as support for one model over another is inviting us to think that the tension ratio, which we can define as
\begin{equation}
T^{AB}_{12} \equiv \frac{R_1^{AB}}{R_2^{AB}} \quad \left[ = \frac{P(D_A,D_B|M_1)P(D_A|M_2)P(D_B|M_2)}{P(D_A,D_B|M_2)P(D_A|M_1)P(D_B|M_2)} \right] \,,
\end{equation}
is akin to the posterior model probability ratio, or rather to the corresponding Bayes' factor in equation (\ref{e:bf}) which is the unambiguously calculable quantity. But is it?

Directly from the definition, equation (\ref{e:rdef}), we can write
\begin{equation}
B_{12}^{AB} = \frac{R_1^{AB}}{R_2^{AB}} \frac{P(D_A|M_1)P(D_B|M_1)}{P(D_A|M_2)P(D_B|M_2)} \,.
\end{equation}
This seems a small step, since it is just taking the ratio of the definition of the tension metric (itself already a ratio) for two different models. But it permits a novel interpretation. Written using Bayes' factors we have
\begin{equation}
\label{e:bayesfactors}
B^{AB}_{12} = T^{AB}_{12} B^A_{12} B^B_{12} \,.
\end{equation}
If we just had dataset $D_A$, we would update the prior model probability ratio by multiplying by $B^A_{12}$, and likewise for $D_B$.

To interpret this, recall the common view of the Bayesian methodology as of repeated updating under new knowledge, where the posterior of one analysis becomes the prior of the next. This works straightforwardly in parameter estimation where the previous posterior probability distribution becomes the new prior.
One might therefore have expected the Bayes' factor of the combined experiments to decompose into the product of those of each experiment. But this same concept does not work straightforwardly for model probabilities, as a posterior model probability does not on its own carry enough information to be further updated by the next experiment --- the parameter distributions that tell us where in parameter space the model fit the data well are also required. Because of this, $B^{AB}_{12} \neq B^A_{12} B^B_{12}$. Equation~(\ref{e:bayesfactors}) shows that the tension ratio is, essentially by definition, an exact quantification of the failure of the update product.

From this equation we now see that the combined dataset can lead to a favouring of (say) $M_2$ over $M_1$ in several ways. $M_2$ could be a better explanation of dataset A, so that $B^A_{12} \ll 1$, or of dataset B, so that $B^B_{12} \ll 1$, or it could reduce the tension relative to $M_1$, so that $R^{AB}_1 \ll R^{AB}_2$. But conversely, we see that a reduction of tension on its own is {\it not} sufficient to demonstrate a preference for $M_2$, because that reduction might be accompanied by a worse Bayes' factor against one or both datasets. Indeed in the particular case of extensions to $\Lambda$CDM such a worsening is rather likely since the extensions necessarily involve a wider parameter space, and hence reduced predictiveness, usually without much improving the best-fit to the individual datasets since $\Lambda$CDM already fits those well.

From this perspective, we see that it requires some level of coincidence for the overall Bayes' factor to be significantly different from unity due to the tension term alone, without the preferred model showing up in either of the datasets independently. This needs the `wrong' model to fit each of the independent datasets as well as does the `right' model, while failing drastically when the two datasets are combined. Moreover, this is `fit' in the model-level sense, requiring equality of the likelihoods averaged over the whole prior model space, not simply equality of the likelihoods of the best fits within each model.

\section{The Hubble tension}

How do our results inform the debate on the Hubble tension? The latest published result from the SH0ES Collaboration \citep{Riess22} places the tension with the {\it Planck} results of \cite{Planck2018par} at the five-sigma level, with Hubble constant $H_0 = (73.0 \pm 1.0) \, {\rm km \, s}^{-1} \, {\rm Mpc}^{-1}$ versus $H_0 = (67.4 \pm 0.5) \, {\rm km \, s}^{-1} \, {\rm Mpc}^{-1}$. This result motivates a sizable fraction of articles currently being submitted to the arXiv cosmology section.

Let's start with the last point above, that for the tension to discriminate amongst models  it would need to be the case that the `wrong' cosmological model fits the separate datasets as well as does the `correct' model. This would not be particularly surprising for data which only constrains the Hubble constant $H_0$, which is essentially an unpredicted free parameter in any cosmological model. This data can hence be perfectly fit by selecting the parameter to coincide with the observed value (note the priors on $H_0$ implied by each model may not always be identical, so the statement is not completely trivial). It would however be a substantial coincidence for the complicated multi-parameter fits required by complex datasets such as the CMB and structure formation.


To see the magnitude of the imbalance for the Hubble tension, we can look at {\it Planck}'s information gain, i.e.\ the compression of the posterior volume versus the prior. It is a staggering $2 \times 10^{14}$ for its six-parameter $\Lambda$CDM fit (Planck 2018 Paper VI \citep{Planck2018par} Table 2, 5th data column), averaging better than a factor of a hundred per parameter direction. Admittedly their prior choices (Table 1 of Planck 2013 XVI \citep{Planck2013par}, with a later adjustment on the perturbation amplitude prior) are very broad, especially on the sound horizon at last scattering. But even the narrower choices made in the similar analysis in the DES Y3 paper \citep{DESyr3}, which are explicitly stated to have been to some extent motivated by knowing the outcome of previous datasets, lead to a compression of $10^7$ (in five parameters, the optical depth not being quoted) when fit to data including {\it Planck}. By contrast, Hubble constant measurements address only one direction in parameter space giving an overall volume compression of around 20.


It is hence {\it a priori} unlikely that new physics would be of such a type as to evade detection in an experiment as constraining as {\it Planck}, yet be revealed when adding the much less constraining Hubble measurement.  Suppose there were indeed new physics, for instance selected randomly from the large set of options provided by the current literature (e.g.\ \citealt{tensionrev,olympics}). As also noted  by \cite{Vag}, considerable care would be required in choosing the `true' parameters of such a model so as to be undetectable by {\it Planck} while generating a tension with $H_0$. But Nature is not choosing its parameters by this criterion. Properly assessing the significance of the putative tension requires it to be seen in that light. This reasoning may also explain why it has proven so difficult to come up with compelling physical models that resolve the tension, with for instance \cite{Keeley} and \cite{Vag2} arguing generically against late-time and early-time solutions, respectively. This is beautifully highlighted in the response function framework of \cite{HVZ}.

Numerous recent articles have demonstrated that various extensions to the $\Lambda$CDM paradigm can reduce the Hubble tension. Note that even any {\it incorrect} extension model is likely to mitigate tension to some extent, because it is unlikely that the tension is minimised by some new parameter taking the value zero (unless that parameter is one-sided). Incorrect models are to be disfavoured because they incur unfavourable Bayes factors against the individual datasets, not because of any effect they have on the tension.

\cite{olympics} provide an extensive comparison of candidate models to relieve the Hubble tension, considering both tension and overall fit. They do not use a fully Bayesian technique, but instead consider three different criteria (ultimately only two are used in practice). The tension is measured by a Gaussian-sigmas or parameter-shift method \citep{Raveri}, and the criterion for model success is a sufficiently reduced tension. The models are then required to pass a test on overall goodness of fit to the data. Unfortunately though, the test uses the Akaike Information Criterion (AIC, \citealt{Akaike, Liddle}) which reflects only the fit quality at the maximum-likelihood point, which is almost guaranteed to improve regardless of the correctness of the chosen model. This test does not capture the reduced predictiveness of extended models embodied in the likelihood averaged over parameter space, which is the quantity we have shown to be the one which offsets gains in tension. 

If, as we are arguing, the evidence for new physics from the Hubble tension is weaker than normally supposed, we must turn to the other explanations. The tension is too great for statistical fluke to be a plausible explanation on its own, though it could be a contributor. Our suspicions therefore rest with systematics, whose situation we briefly summarize for context and trends.

There have long been discrepancies between results from use of different Type Ia supernova light-curve fitters. Indeed in 2010 the WMAP Collaboration's 7-year data release \citep{WMAP7} omitted supernovae entirely from their headline cosmological parameters results due to significant differences in constraints when using the SALT2 \citep{Guy} and MLCS2K2 \citep{Jha} light-curve fitters, even on the same supernova datasets in combination with CMB and other data. To our knowledge this has never been properly resolved, though MLCS2K2 appears to have subsequently fallen out of favour (SH0ES uses the Pantheon+ supernova compilation \citep{Pantheon+}, which uses SALT2). However supernova calibration is less crucial in distance ladder methods for $H_0$ than it is for dark energy measurements.


More recently, several papers have suggested possible sources or symptoms of underestimation of systematic uncertainties. \cite{Efstathiou} discusses the way in which Cepheid sample outliers are removed, and an offset between different geometrical distance indicators and its possible relation to Cepheid metallicity dependence. \cite{Mortsell1} relax the assumption of universal colour--metallicity relations to correct Cepheid magnitudes for dust extinction, maintaining high values of $H_0$ but with somewhat greater uncertainties, while \cite{Mortsell2} investigate several other local distance ladder uncertainties, again mitigating the tension. The impact of colour cuts in the supernova sample was raised by \cite{Wojtak}, though analysis of Pantheon+ (\citealt{Brout22} and A.\ Riess, private communication) strongly limits this. It has also been argued on purely statistical grounds that there may be an unidentified calibration bias \citep{blanchard}, while \cite{Lopez} uses a historical analysis to urge caution. The possible presence of nearby large voids lends a different source of systematic uncertainty to the interpretation \citep{HBK}.

The possibility of underestimated calibration uncertainties is the focus of the recent analysis from the Carnegie Supernova Project \citep{Uddin}, which combines three supernova calibrations: Cepheids, tip of the red giant branch, and surface brightness fluctuations. They argue that the matching of different calibrators implies somewhat larger systematic uncertainties, leading to a modest reduction in the tension (their 
$H$-band calibration result is almost identical to that of SH0ES, while their $B$-band calibration has a central value about one-sigma lower). On the other hand, \cite{Riess23} and \cite{RB23} (see also \citealt{Follin}) have argued that there is no evidence for larger systematics in the Cepheid calibration. \cite{Freedman} provide an additional overview of the current situation.

On the CMB side, discussion of systematics has been more limited, as the nature of the data and its interpretation leaves less to argue about. The systematic differences between the two main likelihood pipelines, Plik and  CamSpec, are well below one-sigma for all parameters and especially for $H_0$. Most concerning at present is the unexpected phenomenological $A_{\rm L}$ parameter \citep{Calabrese}, which gives an indication of an excess of lensing in the CMB temperature power spectrum at more than two-sigma \citep{Planck2018par}. The implications of this for the Hubble tension are described by \cite{tensionrev}.

Finally, one might also take into account the following poll, surveying 59 attendees at the International Astronomical Union (IAU) symposium 376 (held in April 2023) as to the cause of the tension, reported in \cite{RB23}.\footnote{One might say that this is exploiting the Bayesian nature that \cite{Friston} identifies as hard-wired into the brain through the free-energy principle.}
The outcome was 
\begin{center}
\begin{tabular}{lr}
CMB systematics & 5 votes\\ 
$\Lambda$CDM wrong, implying new physics & 36 votes\\ 
Stellar systematics & 18 votes
\end{tabular}
\end{center}
Hence the participants in this poll collectively assessed the significance of the Hubble tension towards new physics as approximately one-sigma.

\section{Conclusion}

We have shown that use of the tension ratio as a proxy for the Bayes factor of the combined data gives an incomplete picture. To interpret the lessened tension of one model versus another as favouring that model, one also needs the Bayes' factors of the individual experiments. But this advice is rather moot in fully Bayesian calculations, because {\it to calculate the tension ratio one has to calculate the combined-data Bayes' factor for each model anyway}, which is the quantity we actually need to perform Bayesian model comparison. We should simply use that to determine which model's position has been improved by the combined data. The tension then reverts to its original purpose, which is to tell us whether, under the favoured model, the datasets are actually compatible or not.

In analyses that are not fully Bayesian, where a statement might be more along the lines of a reduction in tension from $4\sigma$ to $2\sigma$ interpreted via a $p$-value, our arguments still urge caution. It is tempting, but erroneous, to believe that an extra parameter is offering good explanatory value if it reduces dataset tensions by more than one-sigma. Dataset tensions are not the same as goodness-of-fit to the data and nor, as we have seen, are they the whole story. Any analysis that claims evidence for new physics {\it solely} on the basis of alleviating dataset tensions should be considered incomplete and suspect.

There is considerably vast literature on potential `new physics' solutions to the Hubble tension; for example, \cite{Vag2} lists more than 500 papers in this category. This literature does, unfortunately, contain large numbers of papers whose analysis focuses solely on reducing the tension. Even despite that, it has proven alarmingly hard to find well-motivated physical models capable of explaining or even substantially reducing the tension. Further, it is {\it a priori} unlikely that, were such an extended model correct, it would happen to have `true' parameters that so successfully hide their impact on individual datasets yet emerge via dataset tensions. Combining these arguments against the likelihood of a `new physics' explanation for the Hubble tension highlights the importance of continued studies into possible origins via systematics.

\section*{Acknowledgements}
We thank the members of the Dark Energy Survey's Theory and Combined Probes working group members for numerous discussions, particularly Dragan Huterer, Ofer Lahav, Pablo Lemos, and Marco Raveri. We also thank Alan Blanchard, Eoin Colg\'ain, Maria Dainotti, Biagio De Simone, Vasco Gil Gomes, Ariel Goobar, Will Handley, Mustapha Ishak, Alex Kim, Pavel Kroupa, David Parkinson, Adam Riess, Nao Suzuki, and Jes\'us Torrado for discussion and comments, and Andrei Linde for discussion and encouragement. This work was supported by the Funda\c{c}\~{a}o para a Ci\^encia e a Tecnologia (FCT) through the research grants UIDB/04434/2020, UIDP/04434/2020, and PTDC/FIS-AST/0054/2021, and the Investigador FCT Contract Nos.\ CEECIND/02854/2017,  CEECIND/02581/2018 and POPH/FSE (EC). This article/publication is based upon work from COST Action CA21136 -- ``Addressing observational tensions in cosmology with systematics and fundamental physics (CosmoVerse)'', supported by COST (European Cooperation in Science and Technology).
\section*{Data Availability}

 No new data were generated or analysed in support of this research.





\end{document}